\documentclass[aps,twocolumn,floats,nofootinbib,prl]{revtex4}
\usepackage{graphics,graphicx,epsfig}
\usepackage{amssymb,color}
\usepackage{xcolor}
\usepackage{epsf,epstopdf,wrapfig}
\usepackage{amsmath,textgreek,bm}
\usepackage{cancel}
\usepackage[normalem]{ulem} 
\definecolor{green(ryb)}{rgb}{0.4, 0.69, 0.2}

\newcommand{\beq}{\begin{equation}}
\newcommand{\eeq}{\end{equation}}
\newcommand{\beqn}{\begin{eqnarray}}
\newcommand{\eeqn}{\end{eqnarray}}

\begin{document}
\title{Ambiguous signals and efficient codes}

\author{Marianne Bauer$^{a,b}$ and William Bialek,$^{b,c}$}
\affiliation{$^a$Department of Bionanoscience, Kavli Institute of Nanoscience Delft, Technische Universiteit Delft, Van der Maasweg 9, 2629 HZ Delft, the Netherlands\\
$^b$Joseph Henry Laboratories of Physics and Lewis--Sigler Institute for Integrative Genomics,   Princeton University, Princeton, NJ 08544 USA\\
$^c$Initiative for the Theoretical Sciences, The CUNY Graduate Center, 365 Fifth Avenue, New York NY 10016 USA}

\date{\today}

\begin{abstract}
In many biological networks the responses of individual elements are ambiguous.  We consider a scenario in which many sensors respond to a shared signal, each with limited information capacity, and ask that the outputs together convey as much information as possible about an underlying relevant variable.  In a low noise limit where we can make analytic progress, we show that individually ambiguous responses optimize overall information transmission.  
\end{abstract}

\maketitle

The responses of biological systems are more understandable when they point unambiguously to some relevant signal.  Examples include the gene for an enzyme being activated (or de--repressed) in response to increasing concentrations of its substrate \cite{Jacob+Monod_1961,Alberts+al_2002}, or a place cell in the hippocampus becoming active when the animal arrives at a particular location \cite{OKeefe+Dostrovsky_1971,OKeefe+Nadel_1978}.  This simplicity does not mean that the components act alone, but we often expect that even in this more complex setting the responses of individual components have unambiguous meaning.  The violation of this assumption has led, for example, to recent discussions of ``mixed selectivity'' and other non--classical responses in neurons  \cite{Rodgers+al_2021,Tye+al_2024}.  Our goal here is to understand how ambiguous responses can contribute to the optimization of information flow.

The idea that aspects of neural computation serve to optimize information flow emerged not long after the formulation of information theory itself \cite{Barlow_1959,Barlow_1961}, and this continues to guide thinking about neural coding  decades later \cite{Manookin+Rieke_2023}.  Optimization is well posed only if there is a limit on the physical resources that can be deployed,  and perhaps this is even clearer in biochemical and genetic networks where the resources are molecules \cite{tkacik+al_2008a,tkacik+al_2008b}.  It is attractive that the same principles find relevance to biological networks on such different scales \cite{Bialek_2024}. But ambiguity might seem like evidence against optimality.

An extreme example of ambiguous responses is provided by grid cells in mammalian brains, which  respond periodically to the changing position of the animal \cite{Moser+al_2014}.  With many grid cells spanning different periods and phases, the population has enough information to reconstruct position, and it has been argued that this representation is highly efficient \cite{Sreenivasan+Fiete_2011}.  In the developing fruit fly embryo, information about position along the anterior--posterior axis is represented by the concentrations of molecules that are individually ambiguous but precise and unambiguous when taken together \cite{Petkova}, and these are then transformed into spatial patterns that are nearly periodic but with varying phases  \cite{lawrence_1992}.  We are not aiming for a model of these (or other) particular systems, but hope to see such ambiguous encodings emerge from a simple optimization problem, in the spirit of recent work \cite{Bialek_2024,Sokolowski+al_2025,Mijatovic_2025,Zoller_2025}.

\begin{figure}[b]
\centerline{\includegraphics[width = \linewidth]{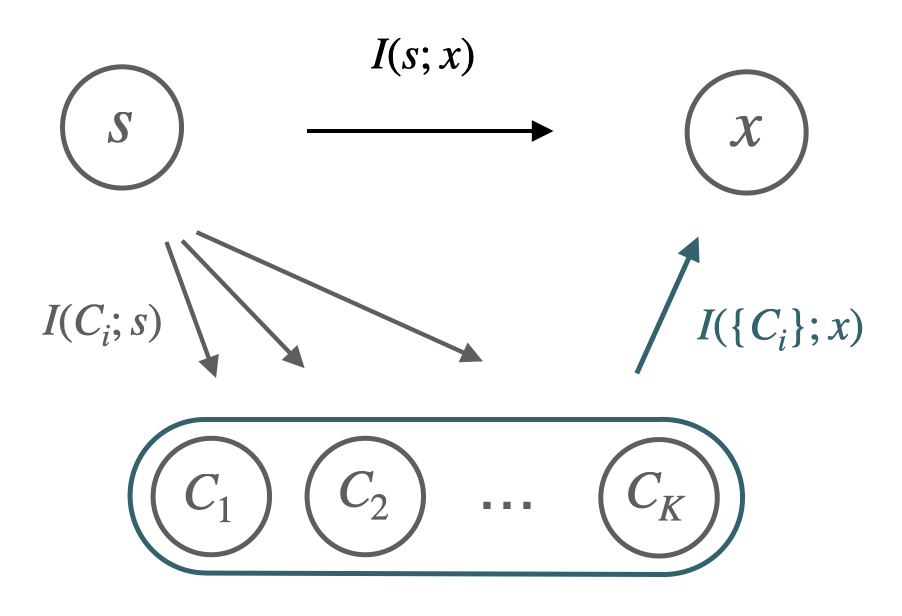}}
	\caption{The information bottleneck problem for multiple variables. A signal $s$ carries information $I(s;x)$ about a relevant variable $x$, which is transmitted via a set of intermediate variables $C_i$. Each $C_i$ is controlled independently by $s$, conveying information $I(C_i;s)$, but they carry information about $x$ collectively, as measured by $I(\{C_i\};x)$ \label{Fig01} }
\end{figure}

We imagine that there is some underlying variable $x$ that is represented initially by a single $s$.  The organism needs to read this signal, but has at its disposal mechanisms with limited information capacity.  It can deploy many of these mechanisms, mapping $s\rightarrow C_i$ for $i=1,\, 2,\, \cdots ,\, K$, but these mappings are independent of one another, schematized in Fig \ref{Fig01}.  The goal is to be sure that despite a limit on the information that can be captured in each of these mappings, the outputs together carry as much information as possible about $x$:  we want to maximize $I(\{C_i\}; x)$ while restricting each $I(C_i; s)$, and this can be done by maximizing
\begin{equation}
{\cal F} = I(\{C_i\}; x) - \sum_{i=1}^K \lambda_i I(C_i; s),
\label{calFdef}
\end{equation}
where $\lambda_i$ is the Lagrange multiplier constraining the information conveyed in the mapping $s\rightarrow C_i$.  Maximizing $\cal F$ is a generalization of the information bottleneck problem \cite{tishby99,Baueretal,bauer2022,bauerbialek23}.   

For simplicity assume that the mapping $\{C_i\} \rightarrow x$ is unambiguous and reasonably precise, although the individual mappings $s \rightarrow C_i$ might be ambiguous.  Concretely we consider
\begin{equation}
P_i(C_i | s) = {1\over{\sqrt{2\pi \sigma_i^2(s)}}}\exp{\bigg\{} - {{[C_i - \bar C_i(s)]^2}\over {2\sigma_i^2 (s)}}{\bigg\}} ,
\end{equation}
where the mean $\bar C_i(s)$ in general is noninvertible. The information
\begin{eqnarray}
I(C_i; s) &\equiv& \int dC \int ds, P_i(C | s) P(s) \log \left[{{P_i(C | s)}\over{P_i(C)}}\right]\\
P_i(C) &=& \int ds, P_i(C | s) P(s)
\end{eqnarray}
then can be written as
\begin{equation}
I(C_i; s) = H[C_i] - {1\over 2}\langle \log[2\pi e \sigma_i^2(s)]\rangle_s ,
\end{equation}
where $\langle \cdots \rangle_s$ denotes an average over the distribution $P(s)$ and as usual the entropy 
$H[C] = -\langle\log P(C)\rangle_C$.

We can also write the mutual information between $\{C_i\}$ and $x$ as
\begin{equation}
I(\{C_i\}; x) = H[x] - \langle H[x|\{C_i\}]\rangle_{\{C_i\}} ,
\end{equation}
where $H[x|\{C_i\}]$ is the entropy of the conditional distribution $P(x|\{C_i\})$. If the map from $x$ to the mean value of $s$ is invertible and has relatively little noise, then we can think of $P(x|\{C_i\}) $ as describing a process by which we estimate $s$ from $\{C_i\}$ and then just propagate this back to $x$.  So if there are small errors $\sigma_s(\{C_i\})$ in the estimate of $s$, we can approximate
\begin{eqnarray}
P(x|\{C_i\}) &\approx& {1\over\sqrt{2\pi\sigma_x^2(\{C_i\})}} \exp{\bigg\{} - {{[x - \bar x (\{C_i\})]]^2}\over 
{2\sigma_x^2(\{C_i\})} }{\bigg\}} \nonumber\\
&&\\
{1\over {\sigma_x^2(\{C_i\})}} &=& {\bigg |} {{d\bar s}\over {dx}}{\bigg |}^2 {1\over {\sigma_s^2(\{C_i\})}} .
\end{eqnarray}
Further, with multiple $C_i$ contributing to the estimate of $s$, in the limit that these estimates are sharp we can write \cite{Bevington+Robinson_2003}
\begin{equation}
{1\over {\sigma_s^2(\{C_i\})}} = \sum_{i,j = 1}^K {{d\bar C_i(s)}\over {ds}}\left[\Sigma^{-1}(x)\right]_{ij} {{d\bar C_j(s)}\over {ds}} ,
\end{equation}
where $\Sigma(x)$ is the covariance of  the $C_i$ at fixed $x$.   

When $s$ is fixed the noise in each $C_i$ is independent, but the fluctuations is $s$ at fixed $x$ are common to all, so we have, again in the small noise approximation,
\begin{eqnarray}
\Sigma_{ij}(x)  &=& \sigma_i (s)\sigma_j (s) \left[ \delta_{ij} + \sigma_s^2 (s) \phi_i(s)\phi_j(s)\right]{\bigg |}_{s = \bar s (x)} \\
\phi_i (s) &=& {1\over{\sigma_i(s)}} {{d\bar C_i(s)}\over {ds}} .
\end{eqnarray}
If we define $\bm{v} = \sigma_s\bm{\phi}$ then we can use
\begin{equation}
\left( \mathbb{I} + \bm{v}\bm{v}^T\right)^{-1} = \mathbb{I} - { {\bm{v}\bm{v}^T} \over{1 + |\bm{v}|^2}} .
\end{equation}
If $\bar s(x)$ is invertible and the noise $\sigma_s(x)$ is small, then we can also relate the entropies
\begin{equation}
H[s] = H[x] + {\bigg\langle}\log  {\bigg |} {{d\bar s(x)}\over{dx}}{\bigg |} {\bigg\rangle}_x .
\end{equation}
Putting the pieces together we find
\begin{eqnarray}
I(\{C_i\}; x) &=& H[s] + \int ds \, P(s) \log\left[{{\Phi(s)}\over{2\pi e}} \right] ,\\
\Phi(s) &=& {{\sum_i \phi_i(s)^2}\over{1 + \sigma_s^2(s) \sum_i\phi_i(s)^2}} ,
\end{eqnarray}
where again in the small noise approximation we can trade dependence on $x$ for dependence on $s$.

In maximizing $\cal F$ we can adjust both the noise levels $\sigma_i(s)$ and the mean responses $\bar C_i(s)$.  Importantly we don't have interactions among the different values of $s$, so we can take derivatives point by point:
\begin{eqnarray}
{{\delta I(\{C_i\};x)}\over{\delta (1/\sigma_{i}^2(s))} } &=& \frac{1}{2} {\bigg |} \frac{d\bar C_i}{ds}{\bigg |}^2 
\left[\frac{1}{\sum_i \phi_{i}^2} - \frac{ \sigma_{s}^2 }{1 + \sigma_{s}^2 \sum_i \phi_{i}^2}\right]   
\label{eq:partial1}\\
{{\delta   I(C_i;s)}\over{\delta (1/\sigma_{i}^2(s))} }  &=& \frac{1}{2} \sigma_{i}^2 .
\label{eq:partial2}
\end{eqnarray}
Then we have the optimization condition
\begin{equation}
{ {\delta {\cal F}}\over{\delta (1/\sigma_{i}^2(s))} } =  0
\Rightarrow {{\delta I(\{C_i\};x)}\over{\delta (1/\sigma_{i}^2(s))} } = \lambda_i {{\delta   I(C_i;s)}\over{\delta (1/\sigma_{i}^2(s))} } .
\end{equation}
Substituting, this becomes
\begin{eqnarray}
\phi_i^2(s)  &=& \frac{1}{\sigma_{s}^2(s)} \frac{\lambda_i}{\Lambda}  \left( \frac{1-  \Lambda}{\Lambda} \right) \label{phi_opt} \\
\Lambda &=& \sum_{j=1}^K \lambda_j
\end{eqnarray}
We recall that $\phi_i(s)$ is the local slope of the (mean) mapping $s\rightarrow C_i$, in units of its noise level, and this result shows that optimizing information transmission means that all these signal-to-noise ratios (SNRs) are determined by the noise in the original mapping $x\rightarrow s$, and different choices of the Lagrange multipliers scale these SNRs in relation to one another.

We can use the result in Eq (\ref{phi_opt}) to rewrite the information that is captured about position,
\begin{equation}
I(\{C_i\}; x) = H[s] - {1\over 2}\langle \log[2\pi e \sigma_s^2 ]\rangle + {1\over 2}\log(1- \Lambda) .
\end{equation}
Again, to leading order in the small noise approximation it doesn't matter whether we are averaging over $x$ or $s$.  We recognize the combination
\begin{equation}
I(s;x) = H[s] - {1\over 2}\langle \log[2\pi e \sigma_s^2(x) ]\rangle_x
\end{equation}
as the mutual information between $s$ and $x$ if the fluctuations in $s$ at fixed $x$ are Gaussian.  Then we have
\begin{equation}
I(\{C_i\}; x) = I(s; x)  + {1\over 2}\log(1- \Lambda) .
\label{ICx}
\end{equation}
This tells us that the information about $x$ captured in the encoding $\{C_i\}$ just misses being all the information that was available from $s$, and the amount that we miss depends on $\Lambda$.  Going back to the definition in Eq (\ref{calFdef}) we see that $\Lambda$ expresses the strength of the constraints on the information capacities of the individual encodings $s\rightarrow C$.  So as $\Lambda \rightarrow 0$ we allow (unrealistically) that each variable $C_i$ can convey an arbitrarily large amount of information, while as $\Lambda \rightarrow 1$ the constraint becomes very tight.  No other terms enter the information $I(\{C_i\}; x)$.

\begin{figure*}[t]
\includegraphics[width = \linewidth]{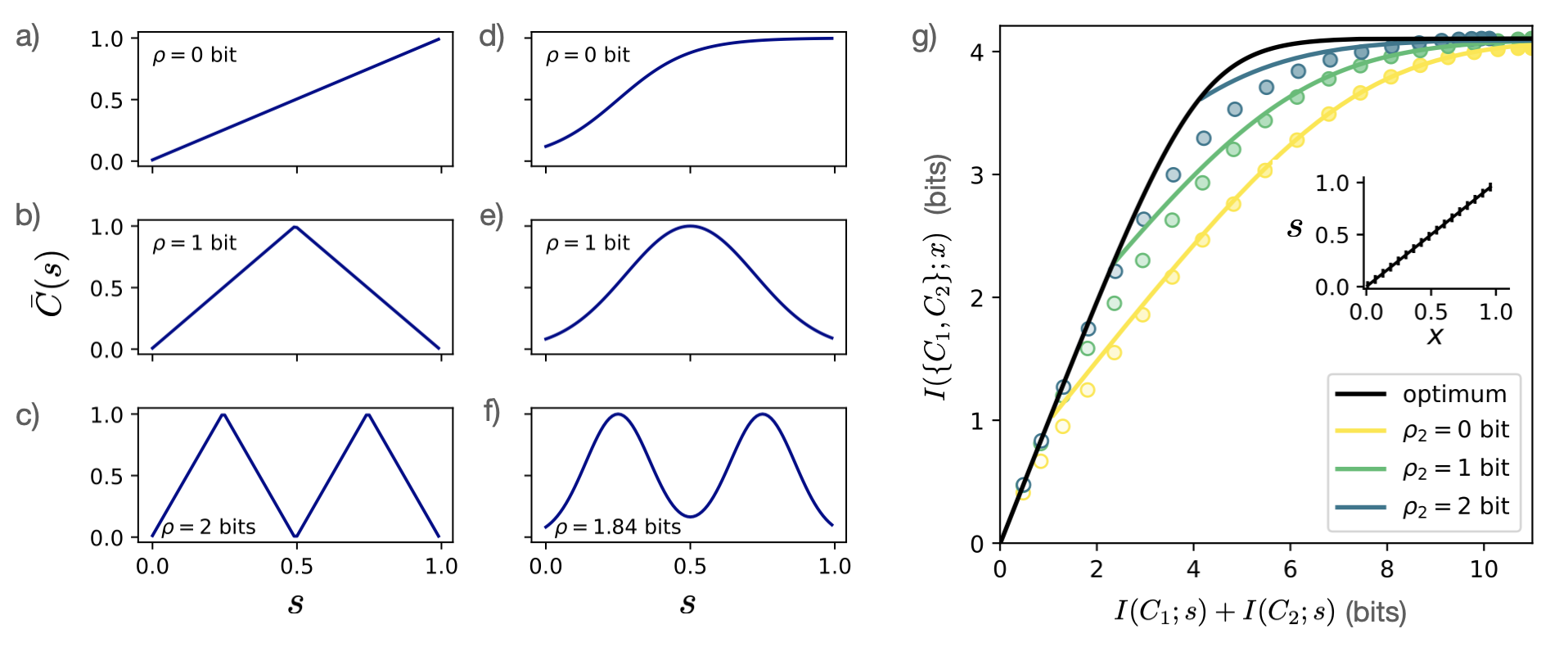}
\caption{Input/output relations $\bar C(s)$, their deficits $\rho$ from Eq~(\ref{rho_def}), and consequences for information flow. 
(a) An invertible relation with $\rho = 0$. 
(b) A sawtooth relation that folds the dynamic range once, with $\rho = 1\,{\rm bit}$. 
(c) A sawtooth with two folds and $\rho = 2\,{\rm bits}$.
(d) A nonlinear relation, still invertible and hence with $\rho =0$ as in (a).
(e) A nonlinear relation with one symmetric fold and  $\rho = 1\,{\rm bit}$ as in (b). 
(f) A double peaked relation, but since the minimum at $s=0.5$ is above the ones at $s=0, 1$ we find a fractional $\rho = 1.84\,{\rm bits}$.  
(g) The information plane for a system in which $s$ encodes the relevant variable $x$ linearly (inset) and information is shared equally between two channels $C_1$ and $C_2$, with $\bar C_1(s)$ being linear and $\bar C_2(s)$ being linear or piecewise linear.  Theoretical predictions (colored lines), numerics (circles), and the optimum when there is a single channel with sufficient capacity (black line).
\label{sawtooth}}
\end{figure*}

Since  $\bar  C_i (s)$ drops out of $I(\{C_i\};x)$ once we optimize the noise levels $\sigma_i(s)$, the only place where they can still enter the optimization problem is through $I(C_i; s)$:
\begin{eqnarray}
 I(C_i;s) &=& 
 H[C_i] - \frac{1}{2} \langle \log \left[ 2 \pi e \sigma_{s}^2\right]\rangle \nonumber\\
 &&\,\,\,\,\, -   {\bigg \langle} \log  {\bigg |} { {d\bar C_i (s) }\over {ds}}{\bigg |} {\bigg \rangle}_s + \frac{1}{2} \log\left( \lambda_i \frac{ 1- \Lambda }{\Lambda^2} \right) \textrm{.}\nonumber\\
 &&
\label{ICs_int}
\end{eqnarray}
We notice that {\em if} the function $\bar C_i (s)$ were invertible, then in the small noise approximation we would have
\begin{equation}
H[C_i] = H[s] +  {\bigg \langle} \log  {\bigg |} { {d\bar C_i (s) }\over {ds}}{\bigg |} {\bigg \rangle}_s .
\end{equation}
In reality the entropy will be smaller, and it useful to define a deficit $\rho$ through
\begin{equation}
H[C_i] = H[s] +  {\bigg \langle} \log  {\bigg |} { {d\bar C_i (s) }\over {ds}}{\bigg |} {\bigg \rangle}_s -\rho_i .
\label{rho_def}
\end{equation}
This allows us to rewrite Eq (\ref{ICs_int}) as
\begin{equation}
 I(C_i;s) = 
 I(s;x) -   \rho_i  + \frac{1}{2} \log\left( \lambda_i \frac{ 1- \Lambda }{\Lambda^2} \right) \textrm{.}
\label{ICs_int}
\end{equation}
We see that $I(C_i;s)$ again is fixed by the Lagrange multipliers, except that it can be reduced by the term $\rho$.

It is useful to consider simple mappings, schematized in Fig \ref{sawtooth}, where we assume that  the signals $s$ are distributed uniformly; note that we have multiple variables $C_i$ and this shows the mean behavior $\bar C_i(s)$ of just one.   If the mapping is invertible we have $\rho = 0$, as in Fig \ref{sawtooth}a and the nonlinear version in Fig \ref{sawtooth}d.  If  the mean response $\bar C(s)$ encodes its input as a sawtooth that ``folds'' the relevant dynamic range symmetrically,  as in Fig \ref{sawtooth}b, then $\rho = \log 2$ or one bit; we get the same answer even if the rise and fall are nonlinear, as in Fig \ref{sawtooth}e.   One bit is substantial on the scale of the information that can flow through a single gene regulatory element \cite{tkacik+al_2008a,tkacik+al_2008b}.  If inputs are folded twice, as in Fig \ref{sawtooth}c, then $\rho = \log 4$, etc..  Nonlinear patterns can give fractional $\rho$, as in Fig \ref{sawtooth}f,  because different values of the response map to different numbers of points on the input axis, and $\rho$ reflects an average over these.

Consider a system that makes equal use of two variables $C_1$ and $C_2$, so that
$I(C_1; s) = I(C_2; s)$, but the first encoding is invertible ($\rho_1 = 0$) and the second is not ($\rho_2 > 0$).
Then with with the relevant information $I(\{C_i\}; x)$ fixed by $\Lambda$ through Eq (\ref{ICx}) the system needs a capacity (in bits)
\begin{eqnarray}
I_{\rm tot} &\equiv& I(C_1; s) + I(C_2;s) \\
& =& I_{\rm tot} (\rho_2 = 0) - \log_2\left({{1+2^{2\rho_2}}\over 2}\right).
\end{eqnarray}
The representation thus is more efficient when information is shared between invertible and ambiguous responses; the effect is bigger for larger $\rho$.

To be more concrete we focus on two variables with $\bar C_1(s)$ being linear (hence invertible) and $\bar C_2(s)$ being  linear or piecewise linear as in Figs \ref{sawtooth}a--c.  For simplicity we assume that $s$ encodes the relevant variable $x$ linearly and that the noise $\sigma_s(x)$ is constant (Fig \ref{sawtooth}g, inset).  From Eq (\ref{phi_opt}) we can see that piecewise linear $\bar C_i(s)$ are solutions only if the $\sigma_i(s)$ also are constant, and their ratio $\sigma_1/\sigma_2$ is set by the condition that information is shared equally between the two channels. With this ansatz we can solve the optimization problem numerically and compare with our theory, which we developed using small noise approximations.  We see in Fig \ref{sawtooth}g that the agreement is good, and better at lower noise levels, as expected. Importantly we confirm that $\bar{C}_2 (s)$ having more folds uniformly improves the transmission of relevant information at fixed capacity, even when corrections to the small noise approximation are noticeable.

None of this would be necessary if a single (invertible) channel had enough capacity to encode all the needed information about $s$ \cite{bauerbialek23}; the optimal performance with this more generous assumption is shown in Fig \ref{sawtooth}g as a black line.  If we need, for example, four bits of information about $s$ in order to have enough information about $x$ and the available mechanisms have a capacity of only two bits, then one might think that dividing the information would be simple, but the different channels typically are redundant with one another; this redundancy is what pushes the colored points/lines to be uniformly below the black line in  Fig \ref{sawtooth}g. 

If one of the channels provides an invertible encoding and the second is folded, then the second channel provides more nearly orthogonal information and the performance moves closer to the bound.   In the limit that there are multiple channels with increasing levels of folding,  the system is (roughly) encoding the different bits in a binary representation of $s$, and these bits are nearly independent of one another.  This is very similar to the argument about the efficiency of grid cell populations as a representation of space in the brain \cite{Sreenivasan+Fiete_2011}.

For all this to make sense, the whole set $\{C_i\}$ must provide an unambiguous encoding of $s$.  Thus folding of the input axis for one of the responses has to be compensated by structure in the other responses. These results  are complementary to previous work on the encoding the expression levels of multiple genes \cite{Baueretal},
where we showed that  the optimal strategy involves single sensors responding to nonseparable combinations of expression levels.

Non--monotonicity and combinatorial coding are essential features of the gap genes and pair--rule genes in the early fly embryo \cite{lawrence_1992}, although the real system is more complicated since the target genes interact and this plays a role in optimizing information transmission \cite{Sokolowski+al_2025}. Stepping back from the example of the fly embryo, there are current discussions about whether non--monotonicity can be achieved by single regulatory elements in or out of equilibrium \cite{Mahdavi+al_2024,Floyd+al_2025,Martinez+al_2025,Andrews+al_2025}, but  it is attractive that we can see these features emerge from general arguments.  In particular, non--monotonicity and combinatorial coding might seem like intrinsically biological elaborations that we would be tempted to ignore in building a physicists' model of the system.  But these elaborations can be seen instead as  solutions to the basic physics problem of transmitting information efficiently through mechanisms with limited information capacity.  Importantly we can reach this conclusion by focusing only on the limited capacity, without having to build microscopic models of the underlying mechanisms.

This work was supported in part by the National Science Foundation through the Center for the Physics of Biological Function (PHY--1734030), by  NWO Talent/VIDI grant NWO/VI.Vidi.223.169, and by the Simons Foundation.

\bibliography{MB+WB_25.bib}
\end{document}